\documentclass[12pt,english]{article}
\usepackage[T1]{fontenc}
\usepackage[latin1]{inputenc}

\makeatletter

 \newcommand{\lyxaddress}[1]{
   \par {\raggedright #1
   \vspace{1.4em}
   \noindent\par}
 }

\usepackage{graphicx}
\usepackage{babel}
\usepackage{setspace}
\usepackage{harvard}
\begin{document}

\doublespace

\title{Low vorticity and small gas expansion in premixed flames}

\author{Bruno Denet$^{1}$ and  Vitaly Bychkov$^{2}$}

\maketitle

\lyxaddress{$^{1}$IRPHE 49 rue Joliot Curie BP 146 Technopole de Chateau
Gombert 13384 Marseille Cedex 13 France}

\lyxaddress{$^{2}$Institute of Physics, Ume\aa\ University, SE--901 87,
Ume\aa, Sweden}

\begin{center}submitted to Combustion Science and Technology\end{center}

\begin{abstract}
Different approaches to the nonlinear dynamics of
premixed flames exist in the literature: equations based on 
developments in a gas expansion parameter, weak nonlinearity approximation,
potential model equation in a coordinate-free form. 
However the relation between these different equations is often unclear. 
Starting here with the low vorticity
approximation  proposed recently by one of the authors, 
we are able to recover from this formulation the dynamical equations 
usually obtained at the lowest orders in
gas expansion for plane on average flames, 
as well as obtain a new second order coordinate-free equation 
extending the potential flow model known as the
Frankel equation.
It is also common to modify gas expansion theories 
into phenomelogical equations, which agree quantitatively 
better with numerical simulations.
We  discuss here what are the restrictions imposed 
by the gas expansion development results on this process.

\end{abstract}


\section{Introduction \label{sec:Introduction}}

The nonlinear description of the Darrieus-Landau (DL) instability
of premixed flames began twenty-five years ago, when Sivashinsky
obtained, as a first-order development in powers of a gas
expansion parameter, a nonlinear equation known today as the
Sivashinsky equation \cite{siva77}. Starting with the first
simulations of Michelson (see for instance \cite{michelsonsiva})
and with the analytical solution of Thual with coauthors
\cite{Thual-et.al-85}, this equation has shown a surprising
qualitative agreement with experiments and direct numerical
simulations
\cite{denethaldenwang,Bychkov-et.al-96,Kadowaki99,Travnikov-et.al-00}.
The only nonlinear term of the Sivashinsky equation has a purely
geometrical origin and is not related to the usual nonlinearity
 of the Navier-Stokes equations. Challenged by Clavin about
the fact that the Navier-Stokes nonlinearity should induce
modifications of the flame equation, particularly for realistic
gas expansion, Sivashinsky went on to show, in a joint paper with
Clavin \cite{sivaclavin}, that even at the second order in gas
expansion, the equation obtained was still an equation with the
same terms, only with modified coefficients. Today various methods
try to improve on these equations by using the approximation of
weak nonlinearity or next orders in gas expansion
\cite{ZhdanovTrubnikov,Joulin91,Bychkov98,kazakovcst2002,boury}.
On the other hand, similar to the original Sivashinsky equation,
all these equations have been derived for the planar on average
flame front, and are valid only when the slope is not too large.
Actually, different variations of the Sivashinsky equation have
been constructed for different geometries like expanding flames
\cite{filyandsiva,dangelojoulinboury} or oblique flames
\cite{bouryjoulin.ctm2002}, when the flame shape  departs slightly
from the unperturbed case.

A different equation was introduced in the theoretical community
by Frankel \cite{frankel} in 1990, although parts of this approach
were anticipated by  numerical studies, see for instance
\cite{ghoniemchorinoppenheim}. The
Frankel equation was derived in a coordinate-free form (in two or
three dimensions) using similarities between electrostatics and
the flame-generated flow. Constructing his theory Frankel assumed
 that the flow is potential everywhere and neglected vorticity generated by
the flame in the burnt gas. As a matter of fact, such an
assumption originated in the analysis by Sivashinsky
\cite{siva77}. Numerical simulations demonstrated that the Frankel
equation describes qualitatively well the nonlinear evolution of
expanding flames
\cite{frankelsiva,blinnikovsasorov,ashurst97,denetfrankel}  and
 oblique flames  \cite{denetbunsen2d} (even
recently in three dimensions \cite{denetbunsen3d}). The Frankel
equation became especially popular in the studies of fractal
flames developing because of the DL instability on large scales
\cite{blinnikovsasorov,denetfrankel}.

 From a theoretical point of view, the relationship between the
Frankel and Sivashinsky equations was put forward from the very
beginning, since the original  paper \cite{frankel} showed that
the Frankel equation reduces to the Sivashinsky equation for plane
on average flames (with lateral boundaries at infinity) and for expanding
flames. However, validity of the Frankel equation has been
questioned rather often because of the assumption of the potential
flow, which, in principle, violates hydrodynamic conservation laws
of the flame front. The original analysis \cite{frankel} did not
demonstrate if the Frankel equation follows from the hydrodynamic
equations in the limit of small gas expansion, as it was done for
the Sivashinsky equation \cite{siva77}.

Recently, however, one of us \cite{bychkovzaytsevakkerman}
reconsidered the problem and introduced a low vorticity
approximation (compared to the Frankel case, where the vorticity
is strictly zero), which was justified by the previous analysis of
curved flames \cite{Bychkov98}. The approximation of low vorticity
enabled  to derive a system of coordinate-free equations
describing the evolution of the front even for realistically large
thermal expansion of the burning matter. This system of equations
is rather complex and has not been successfully solved numerically
for the moment. In the present paper, we develop this system up to
the second order in gas expansion, which  is equivalent to
developing the hydrodynamical equations at this order. Such
calculation proves the Frankel equation at the first order in gas
expansion, and leads to a second order form of the  equation (we
will however insist on some difficulties specific to the oblique
flame geometry). This form in turn can be demonstrated to contain
the Sivashinsky-Clavin equation in the planar on average case.

In Section \ref{sec:derivation}
 we obtain this second order form.
In Section \ref{sec:Reduction}
 this equation will be reduced to
the Sivashinsky-Clavin equation. In Section
\ref{sec:Time-derivative}
 we discuss some basic problems related
to the Sivashinsky-Clavin equation and expansion in powers of a
small parameter in general. Finally, Section  \ref{sec:Conclusion}
contains a conclusion.

\section{Derivation of the second order Frankel equation \label{sec:derivation}}

Let us start this section by summarizing the main points of the
low vorticity approximation of \cite{bychkovzaytsevakkerman} (the
reader is invited to read this article for more details). This
approximation is derived from the hydrodynamical equations :

\[
\mathbf{\nabla\cdot u}=0,\]

\[
\frac{\partial\mathbf{u}}{\partial\tau}+\left(\mathbf{u\cdot
\nabla}\right)\mathbf{u}=-\vartheta\mathbf{\nabla}\Pi,\]
 where the
equations written are made non-dimensional with the use of the
laminar flame velocity $U_{f}$ and a reference length $R$. The
non-dimensional velocity is noted $\mathbf{u}$ and the pressure
$\Pi=\left(P-P_{f}\right)/\rho_{f}U_{f}^{2}$, $\vartheta=1$ in the
fresh mixtures and $\vartheta=\Theta$ in the burnt gases. The
flame is considered as a discontinuity, $\Theta=\rho_{f}/\rho_{b}$
is the ratio of density in fresh  and burnt gases. The parameter
$\left(\Theta-1\right)$ will be the parameter of the expansion.
The boundary conditions of the flame can be classically shown to
lead to
\[
\mathbf{u}_{+}=\mathbf{u}_{-}+\left(\Theta-1\right)\mathbf{n},\]
\[
\Pi_{+}=\Pi_{-}+1-\Theta,\]
 where $\mathbf{n}$ is the normal vector
to the flame surface, directed towards the burnt gases. The first
boundary condition accounts for both the jump of normal velocity
and conservation of tangential velocity at the front. We introduce
the velocity potentials in fresh ($-$) and burnt ($+$) matter
$\phi_{\pm}$, which satisfy the Laplace equation and are defined
by
\[
\mathbf{u}_{-}=\mathbf{\nabla}\phi_{-},\]
\begin{equation}
\mathbf{u}_{+}=\mathbf{u}_{p+}+\mathbf{u}_{\upsilon+}=\mathbf{\nabla}\phi_{+}+
\mathbf{u}_{\upsilon+},
\label{eq:separation}\end{equation}
where
$\mathbf{u}_{p+}$ and $\mathbf{u}_{\upsilon+}$ are the potential
and vortical parts of the velocity field in the burnt gases.

We work in a local system of coordinates moving with the flame
front at the velocity $-V_{s}\mathbf{n}$ such as
$\mathbf{\nabla}_{s}=\mathbf{e}_{t}\cdot\mathbf{\nabla}$, where
$\mathbf{e}_{t}$ is the unit tangential vector, and

\[
\frac{\partial}{\partial\tau_{s}}=\frac{\partial}{\partial\tau}-V_{s}\frac{\partial}{\partial\mathbf{n}}.\]
Using basic properties of Green functions of the Laplace equation,
Bernoulli integrals, and boundary conditions at the front, the
following system of equations (low vorticity limit) is obtained
\begin{equation}
\frac{\partial}{\partial\tau_{s}}\left(\phi_{+}-\phi_{-}+
\phi_{-}\left(1-\Theta\right)\right)=\frac{\Theta-1}{2}u_{-}^{2}-
\left(\Theta-1\right)V_{s}^{2}+\left(\Theta-1\right)V_{s}+
V_{s}u_{\upsilon n+}+f,
 \label{eq:lowvorticity_eq1}
 \end{equation}
\begin{equation}
\frac{\partial\mathbf{u}_{\upsilon}}{\partial\tau}+\left(\mathbf{u}_{p+}\cdot
\mathbf{\nabla}\right)\mathbf{u}_{\upsilon}=0.
\label{eq:lowvorticity_intermediate}
\end{equation}
Note that the form given here is not the final form of
\cite{bychkovzaytsevakkerman}, here we do not incorporate $\Theta$
into the variables in order to make the development in
$\left(\Theta-1\right)$ easier. Equation
(\ref{eq:lowvorticity_eq1}) comes from the Bernoulli integral; $f$
is generally a function of time appearing in the Bernoulli
integrals (unlike \cite{bychkovzaytsevakkerman} we have included a
constant term $\left(\Theta-1\right)^{2}/2$ into $f$). We will
choose however not to include any additive terms containing a
function of time in the potentials, so that $f$ has to be
considered as a constant. Furthermore, strictly speaking, we could
add a constant in $\mathbf{u}_{\upsilon+}$ and subtract it from
$\mathbf{u}_{p+}$ but naturally $\mathbf{u}_{\upsilon+}$ has to be
taken as small as possible, which makes the Bernoulli integral a
good approximation of the Navier-Stokes equation in the burnt
gases. A non-zero value of $f$ would lead to a constant value of
$u_{\upsilon n+}$ at infinity (convected by the potential
velocity) for a plane or oblique geometry, so that we must have in
this case $f=0$. Similarly, in the case of the expanding geometry,
a non zero value of $f$ would leave  a constant value of
$u_{\upsilon n+}$ behind the front, which is not possible because
there is no source or sink present inside the expanding flame.

Equation (\ref{eq:lowvorticity_intermediate}) expresses the fact
that the vortical part of the velocity field is convected by the
potential part (it will create a shear flow at infinity in the
plane configuration that would disappear far from the flame if
viscosity was included). Note that this equation is different from
the equivalent one given in \cite{bychkovzaytsevakkerman}, which
included only the time derivative. As formulated in
\cite{bychkovzaytsevakkerman}, the approach of low vorticity takes
into account only convection by the uniform component of the
velocity field. In the geometry of planar (on average) flame front
propagating along z-axis this corresponds to the drift term
$\Theta {\partial \mathbf{u}_{\upsilon}}/ {\partial z} $. In the
 case of expanding flames there is no uniform velocity in the
burnt matter and the vorticity created at the flame surface is
simply left behind as the flame radius increases; by this reason
the drift term was omitted in \cite{bychkovzaytsevakkerman}. In
the present paper we are interested in small gas expansion of the
flame equations, which allows to consider a more general form of
equation (\ref{eq:lowvorticity_intermediate}). Note also that at
the dominating
 order in gas expansion, the convection by the potential velocity
  is simply equivalent to the
 convection by the injection velocity. Developing equation
(\ref{eq:lowvorticity_intermediate}) and using the continuity
equation ($n$ is the normal, $s$ represents the tangential
coordinates)
\[
\frac{\partial u_{\upsilon n+}}{\partial n}+\nabla_{s}\cdot
\mathbf{u}_{\upsilon t+}=0\] and $\mathbf{u}_{\upsilon
t+}=\nabla_{s}\left(\phi_{-}-\phi_{+}\right)$, we obtain
\begin{equation}
\frac{\partial\mathbf{u}_{\upsilon}}{\partial\tau_{s}}\cdot
\mathbf{n}+\left(\left(\mathbf{u}_{pt+}\cdot
\mathbf{\bigtriangledown}_{s}\right)\mathbf{u}_{\upsilon}\right)\cdot
\mathbf{n}=\left(V_{s}+u_{pn+}\right)\nabla_{s}^{2}\left(\phi_{-}-\phi_{+}\right),
\label{eq:lowvorticity_eq2}
\end{equation}
 where $V_{s}=1-u_{n-}$.
Compared to \cite{bychkovzaytsevakkerman}, we have two
supplementary terms coming from the convection by the potential
flow field. The first term of equation (\ref{eq:lowvorticity_eq2})
(the term with the time derivative) is also slightly corrected
compared to this paper.

The Laplace equation leads to an equation with a different form in
two and three dimensions:
\begin{equation}
3D:\qquad\phi_{+}+\phi_{-}=-\frac{1}{2\pi}\int\left(\frac{\Theta-1-
u_{\upsilon n+}}{\left|\mathbf{r}_{s}-\mathbf{r}\right|}+
\left(\phi_{+}-\phi_{-}\right)\mathbf{n}\cdot
\frac{\mathbf{r}_{s}-\mathbf{r}}{\left|\mathbf{r}_{s}-\mathbf{r}\right|^{3}}\right)dS
(\mathbf{r}_{s}),
 \label{eq:green_3d}
\end{equation}
or
\begin{equation}
2D:\qquad\phi_{+}+\phi_{-}=-\frac{1}{\pi}\int\left(\left(\Theta-1-
u_{\upsilon n+}\right)\ln\left|\mathbf{r}_{s}-\mathbf{r}\right|-
\left(\phi_{+}-\phi_{-}\right)\mathbf{n}\cdot
\frac{\mathbf{r}_{s}-\mathbf{r}}{\left|\mathbf{r}_{s}-
\mathbf{r}\right|^{2}}\right)dl(\mathbf{r}_{s}).
\label{eq:green_2d}
\end{equation}
Now let us expand all variables in powers of
$\left(\Theta-1\right)$:

$\phi_{\pm}=\phi_{\pm}^{(1)}+\phi_{\pm}^{(2)}+\cdots$

$u_{-}=u_{-}^{(1)}+u_{-}^{(2)}+\cdots$

$V_{s}=V_{s}^{(0)}+V_{s}^{(1)}+V_{s}^{(2)}\cdots$

$u_{\upsilon n+}=u_{\upsilon n+}^{(1)}+u_{\upsilon
n+}^{(2)}+\cdots$

$\frac{\partial}{\partial\tau_{s}}=O\left(\Theta-1\right)$,

where the subscript $^{(i)}$ means that the term is of order
$\left(\Theta-1\right)^{i}$. Note also that we do not expand the
positions $\mathbf{r}$ in powers of $\left(\Theta-1\right)$; this
can be done only in particular geometries, when   the difference
between the actual and the unperturbed positions is of order
$O\left(\Theta-1\right)$. To reduce the equation obtained to the
Sivashinsky-Clavin equation in the planar geometry, we will have
to consider the fact (see section \ref{sec:Reduction}) that the
vertical coordinate is of order $\left(\Theta-1\right)$, but this
order of magnitude estimate is not geometry-independent.

First, using the relation $V_{s}=1-u_{n-}$ we find that
$V_{s}^{0}=1$. Note that at this zeroth order, we can have a term
$\mathbf{V}_{boundary}^{(0)}\equiv\mathbf{V}_{inj}$ (a constant
injection velocity, for instance) that has to be added to the
velocity field in order to satisfy the boundary conditions at
infinity. At each order, we will encounter a
$\mathbf{V}_{boundary}$ term, so let us define it properly.
$\mathbf{V}_{boundary}$ is a velocity field obeying the Laplace
equation, without jumps on the flame surface, which helps
satisfying  the boundary conditions at infinity. Although we will
perform the calculations in a reference frame without injection
velocity, let us note that with an injection velocity we would
have $V_{s}^{0}=1-V_{injn}$. Obviously, the Bernoulli relation is
Galilean invariant, so we choose the reference frame where the
calculations are simpler, knowing that at the end, the injection
velocity (if present) may be added to the final formula.

By developing equation (\ref{eq:lowvorticity_eq1}) we have at the
$O(1)$ order
\[
-\left(\Theta-1\right)V_{s}^{(0)2}+\left(\Theta-1\right)V_{s}^{(0)}+V_{s}^{(0)}u_{\upsilon
n+}^{(1)}=0,\]
which, with $V_{s}^{(0)}=1$, leads to $u_{\upsilon
n+}^{(1)}=0$. We also have, according to equation
(\ref{eq:lowvorticity_eq2})
\[
0=\nabla_{s}^{2}\left(\phi_{-}^{(1)}-\phi_{+}^{(1)}\right),\]
which implies in three dimensions, using equation
(\ref{eq:green_3d}) and the value of $u_{\upsilon n+}^{(1)}$
\begin{equation}
\phi_{-}^{(1)}=\phi_{+}^{(1)}
=-\frac{1}{4\pi}\int\frac{\Theta-1}{\left|\mathbf{r}_{s}-\mathbf{r}\right|}dS
(\mathbf{r}_{s}).
\label{eq:phi_frankel_3d}
\end{equation}
This is
the potential of a uniformly charged surface. Apart from
multiplicative factors we will call $\left(\Theta-1\right)$ the
surface charge (or line charge for the corresponding two
dimensional equation). The reader is referred to
\cite{denetbunsen2d} for a simple presentation of this
electrostatic analogy. This potential leads to the corresponding
velocity
\begin{equation}
V_{s}^{(1)}=-u_{n-}^{(1)}=\frac{\left(\Theta-1\right)}{2}
\left(1+\frac{1}{2\pi}\int\frac{\mathbf{n}\cdot
(\mathbf{r_{s}}-\mathbf{r})}{\left|\mathbf{r}_{s}-
\mathbf{r}\right|^{3}}dS(\mathbf{r}_{s})\right)-\mathbf{V}_{boundary}^{(1)}\cdot
\mathbf{n}.
\label{eq:frankel}
\end{equation}
Up to the first order, $V_{s}=1+V_{s}^{(1)}$, which is exactly the
Frankel equation in three dimensions (when the flame front is a
surface).

Now let us consider the $O(2)$ terms. At this order, we have from
equation (\ref{eq:lowvorticity_eq1})
\[
-2\left(\Theta-1\right)V_{s}^{(0)}V_{s}^{(1)}+\left(\Theta-1\right)V_{s}^{(1)}+V_{s}^{(0)}u_{\upsilon
n+}^{(2)}+V_{s}^{(1)}u_{\upsilon n+}^{(1)}=0.\]
With the previous
computed values $V_{s}^{(0)}=1$ and $u_{\upsilon n+}^{(1)}=0$, it
reduces to
\begin{equation}
u_{\upsilon
n+}^{(2)}=\left(\Theta-1\right)V_{s}^{(1)}.
\label{eq:vorticalvelocity}
\end{equation}
Taking into account that $\phi_{-}^{(1)}=\phi_{+}^{(1)}$ equation
(\ref{eq:lowvorticity_eq2}) gives
\begin{equation}
\nabla_{s}^{2} \left( \phi_{-}^{(2)}-\phi_{+}^{(2)}
\right)=\left[\left(\mathbf{u}_{inj}\cdot
\mathbf{\bigtriangledown}_{s}\right)\mathbf{u}_{\upsilon}^{(2)}\right]\cdot
\mathbf{n}, \label{eq:phi_2}
\end{equation}
so we obtain from
(\ref{eq:green_3d}) in three dimensions with the previous value of
$u_{\upsilon n+}^{(2)}$

\[
\phi_{\pm}^{(2)}=-\frac{1}{4\pi}\int\left(\frac{-\left(\Theta-1\right)V_{s}^{(1)}}
{\left|\mathbf{r}_{s}-\mathbf{r}\right|}-\left(\phi_{-}^{(2)}-
\phi_{+}^{(2)}\right)\mathbf{n}\cdot\frac{\mathbf{r}_{s}-\mathbf{r}}{\left|\mathbf{r}_{s}
-\mathbf{r}\right|^{3}}\right)dS(\mathbf{r}_{s})\mp\left(\phi_{-}^{(2)}-\phi_{+}^{(2)}\right)/2\]
with $\left(\phi_{-}^{(2)}-\phi_{+}^{(2)}\right)$ determined by
equation (\ref{eq:phi_2}). Let us note however that this term
exists only when there is a strong tangential velocity field (i.e.
for oblique flames). In the plane and expanding flame cases, the
tangential velocity field can be neglected, so that
$\left(\phi_{-}^{(2)}-\phi_{+}^{(2)}\right)$ is also negligible in
the previous formula. In the rest of the article, we will only
write the formulas with this term neglected, but let us insist on
the fact that by doing so, we neglect a dipolar contribution to
the potential, which could be important in oblique cases.

Summing the first and second order terms, we have
\begin{equation}
\phi_{-}^{(1)}+\phi_{-}^{(2)}=\phi_{+}^{(1)}+\phi_{+}^{(2)}=
-\frac{1}{4\pi}\int\frac{\left(\Theta-1\right)-
\left(\Theta-1\right)V_{s}^{(1)}}{\left|\mathbf{r}_{s}-
\mathbf{r}\right|}dS(\mathbf{r}_{s}),
\label{eq:phi_total_3d}
\end{equation}
which can be interpreted as a local surface charge modified from
$\Theta-1$ to
\begin{equation}
\Theta-1-\left(\Theta-1\right)V_{s}^{(1)}.
\label{eq:charge}
\end{equation}
Then the total velocity field up to the second order is
\begin{eqnarray}
V_{s}^{(0)}+V_{s}^{(1)}+V_{s}^{(2)} & \textrm{=} & 1-u_{n-}^{(1)}-u_{n-}^{(2)}\nonumber \\
 & = & 1+\frac{\Theta-1-\left(\Theta-
 1\right)V_{s}^{(1)}}{2}\left(1+\frac{1}{2\pi}\int\frac{\mathbf{n}\cdot
 (\mathbf{r_{s}}-\mathbf{r})}{\left|\mathbf{r}_{s}-\mathbf{r}\right|^{3}}dS(\mathbf{r}_{s})\right)\nonumber \\
 &  & -\mathbf{V}_{boundary}^{(0)}\cdot \mathbf{n}-\mathbf{V}_{boundary}^{(1)}\cdot
 \mathbf{n}-\mathbf{V}_{boundary}^{(2)}\cdot \mathbf{n},
 \label{eq:frankel_2_3d}
 \end{eqnarray}
or using $V_{s}^{(1)}=V_{s}-1+O(\Theta-1)$ we find the equivalent formulation
\begin{eqnarray}
 V_{s}= 1+\frac{\Theta-1}{2}(2-V_{s})\left(1+\frac{1}{2\pi}\int\frac{\mathbf{n}\cdot
 (\mathbf{r_{s}}-\mathbf{r})}{\left|\mathbf{r}_{s}-\mathbf{r}\right|^{3}}
 dS(\mathbf{r}_{s})\right)-\mathbf{V}_{boundary}\cdot \mathbf{n}
 .
 \label{eq:frankel_2_3d-version}
 \end{eqnarray}
Formula (\ref{eq:frankel_2_3d}) is the coordinate-free
equation for the flame front velocity obtained within the second
order in gas expansion; the first order gives the Frankel
equation. We recall that $V_{s}$ is the normal velocity of the
front propagation, including laminar flame speed and induced
velocity field. Curvature and strain effects are not included, see
\cite{bychkovzaytsevakkerman}. In two dimensions, the equation,
obtained by similar calculations is
\begin{eqnarray}
V_{s}^{(0)}+V_{s}^{(1)}+V_{s}^{(2)} & \textrm{=} & 1-u_{n-}^{(1)}-u_{n-}^{(2)}\nonumber \\
 & = & 1+\frac{\Theta-1-\left(\Theta-1\right)V_{s}^{(1)}}{2}\left(1+\frac{1}{\pi}\int\frac{\mathbf{n}
 \cdot (\mathbf{r_{s}}-\mathbf{r})}{\left|\mathbf{r}_{s}-
 \mathbf{r}\right|^{2}}dl(\mathbf{r}_{s})\right)\nonumber \\
 &  & -\mathbf{V}_{boundary}^{(0)}\cdot \mathbf{n}-
 \mathbf{V}_{boundary}^{(1)}\cdot
 \mathbf{n}-\mathbf{V}_{boundary}^{(2)}\cdot\mathbf{n},
 \label{eq:frankel_2_2d}
 \end{eqnarray}
 or
\begin{eqnarray}
 V_{s}= 1+\frac{\Theta-1}{2}(2-V_{s})\left(1+\frac{1}{\pi}\int\frac{\mathbf{n}
 \cdot (\mathbf{r_{s}}-\mathbf{r})}{\left|\mathbf{r}_{s}-
 \mathbf{r}\right|^{2}}dl(\mathbf{r}_{s})\right)-\mathbf{V}_{boundary}\cdot \mathbf{n}
 .
 \label{eq:frankel_2_2d-version}
 \end{eqnarray}
In order to solve  this equation numerically, the front would have
to be described with marker particles moving according to the
equation
\begin{equation}
\frac{d\mathbf{r}}{dt}=-V_{s}\mathbf{n}.
 \label{eq:propagation}
\end{equation}
Numerical solution to equation (\ref{eq:propagation}) may involve
reconnections in two and three dimensions; possible ways to
overcome these difficulties are described in
\cite{denetbunsen2d,denetbunsen3d}.

\section{Reduction to the Sivashinsky-Clavin equation \label{sec:Reduction}}

We have previously said that equation (\ref{eq:frankel_2_2d})
obtained in the coordinate-free case is supposed to be the
equivalent  of the Sivashinsky-Clavin equation in the plane case.
Indeed, this last equation is derived by a development which is
one order higher than the Sivashinsky equation, while equation
(\ref{eq:frankel_2_2d}) is one order higher than the Frankel
equation. However, transition from one case to the other is not
obvious at all. The Sivashinsky-Clavin equation contains the same
terms with different coefficients, but our modified Frankel
equation contains a surface (or line) charge, which is not
apparently constant at every position on the front. Furthermore,
the order of magnitude of the flame front position makes another
problem, since the position is supposed to be of order
$O\left(\Theta-1\right)$ for the planar case and $O(1)$ in the
coordinate-free case. So, is it possible to recover the
Sivashinsky-Clavin equation from equation (\ref{eq:frankel_2_2d})?
Naturally, we have the help of the original Frankel article
\cite{frankel}, where it was shown for the planar (on average)
case with lateral boundaries at infinity and for the circular
expanding case that the Frankel equation reduces to the
Sivashinsky equation. Let us start by repeating this reasoning
before going to the next order. In two dimensions, the Frankel
equation gives:

\[
V_{s}=1+\frac{\Theta-1}{2}\left(1+\frac{1}{\pi}\int\frac{\mathbf{n}\cdot
(\mathbf{r_{s}}-\mathbf{r})}{\left|\mathbf{r}_{s}-\mathbf{r}\right|^{2}}dl(\mathbf{r}_{s})\right)-
\mathbf{V}_{boundary}^{(0)}\cdot
\mathbf{n}-\mathbf{V}_{boundary}^{(1)}\cdot \mathbf{n}.\] In the
planar case, the normal vector is

\[
\mathbf{n}(\mathbf{r})=\left[-\alpha_{y},1\right]/\sqrt{1+\alpha_{y}^{2}},\]
where $y$ is the lateral coordinate, $\alpha$ is the vertical
position of the front, the vertical $z$ coordinate is positive
towards the burnt gases, $\alpha_{y}$ is a notation for the $y$
derivative. Then equation (\ref{eq:propagation}) gives

\begin{equation}
\frac{\alpha_{t}}{\sqrt{1+\alpha_{y}^{2}}}=-V_{s}.
\label{eq:cos_project}
\end{equation}
In order to recover the Sivashinsky equation, we have to develop
the square root, so we suppose $\alpha=O\left(\Theta-1\right)$ and
$y=O(1)$. We have

\[
\alpha_{t}=-1-\frac{\alpha_{y}^{2}}{2}-\frac{\Theta-1}{2}+\frac{\Theta-1}{2\pi}\,
p.v.\int_{-\infty}^{+\infty}\frac{\alpha\left(\chi+y\right)
-\alpha\left(y\right)-\alpha_{y}\left(y\right)\chi}{\chi^{2}}d\chi+V_{z\,
boundary}^{(0)}+V_{z\, boundary}^{(1)}.\]
Frankel has shown that
the principal value in this formula is

\[
\frac{\Theta-1}{2}I\left(\alpha\right),\]
where
$I\left(\alpha\right)$ is the Landau operator (multiplication by
$\left|k\right|$ in Fourier space, see \cite{frankel} for
details).

Let us take as usual the hypothesis that the injection velocity is
parallel to the $z$ direction and has the value
$\mathbf{V}_{boundary}^{(0)}=\mathbf{V}_{inj}=\left[0,1\right]$.
This term simplifies itself with the $-1$ of the previous
equation. All the following terms of the boundary velocity are
obtained by saying that they do not change the injection velocity.
If this velocity is imposed at a finite distance, then this is
simply obtained by the same integral as in the Frankel equation,
with the same charge, but for the mirror image of the front
relative to the injection location (see \cite{denetbunsen2d} for
an example). If the injection is moved to infinity, then the
velocity field, according to the Gauss theorem, will be the same
as the velocity generated by a plane front, but with a charge
multiplied by the ratio of the front surface to the surface of an
equivalent plane front $V_{z\,
boundary}^{(1)}=\frac{\Theta-1}{2}\left\langle
\sqrt{1+\alpha_{y}^{2}}\right\rangle =\frac{\Theta-1}{2}$, where
$\left\langle \right\rangle $ is a lateral mean value. Higher
orders will be obtained by developing the square root and
replacing $\Theta-1$ by
$\Theta-1-\left(\Theta-1\right)V_{s}^{(1)}$, as discussed in the
previous section. For the time being however, the boundary term
only leads to suppress the term $-\left(\Theta-1\right)/2$ . We
thus obtain the Sivashinsky equation (without curvature terms)
\begin{equation}
\alpha_{t}+\frac{\alpha_{y}^{2}}{2}=\frac{\Theta-1}{2}I\left(\alpha\right).
\label{eq:Sivashinsky}
\end{equation}
Note that with $\alpha=O\left(\Theta-1\right)$ and
$\frac{\partial}{\partial t}=O\left(\Theta-1\right)$ all the terms
of the equation are of the second order.

Let us now perform the same calculation at the next order starting
from the two-dimensional equation (\ref{eq:frankel_2_2d})
\[
V_{s}=1+\frac{\Theta-1-\left(\Theta-1\right)V_{s}^{(1)}}{2}\left(1+
\frac{1}{\pi}\int\frac{\mathbf{n}\cdot(\mathbf{r_{s}}-
\mathbf{r})}{\left|\mathbf{r}_{s}-\mathbf{r}\right|^{2}}dl(\mathbf{r}_{s})\right).\]
Equation (\ref{eq:cos_project}) is still valid. We have, retaining
terms up to the third order
\[
\alpha_{t}=-1-\frac{\alpha_{y}^{2}}{2}-\frac{\Theta-1-
\left(\Theta-1\right)V_{s}^{(1)}}{2}-\frac{\Theta-1}{4}\alpha_{y}^{2}
+\frac{\Theta-1-\left(\Theta-1\right)V_{s}^{(1)}}{2}I\left(\alpha\right).\]
As before, the constant terms (not depending on $y$)  are
eliminated by introducing the boundary velocity terms, but what do
we do about $V_{s}^{(1)}$, given in equation (\ref{eq:frankel}),
which depends on position? The answer is simple: since now we have
 $\alpha=O\left(\Theta-1\right)$, then  we obtain
a product of $\Theta-1$ and $\alpha$ in $V_{s}^{(1)}$. This term
is not
 of the first order anymore, it is now of the second order and can be
neglected (note that this is not necessarily the case in all
geometries). The constant terms in $V_{s}^{(1)}$ are as usual
suppressed by the boundary velocity, and we find
\[
\alpha_{t}=-1-\frac{\alpha_{y}^{2}}{2}-\frac{\Theta-1
-\left(\Theta-1\right)\left(\left(\Theta-1\right)\frac{I\left(\alpha\right)}{2}\right)}{2}
-\frac{\Theta-1}{4}\alpha_{y}^{2}+\frac{\Theta-1-\left(\Theta-1\right)\left(\left(\Theta-
1\right)\frac{I\left(\alpha\right)}{2}\right)}{2}I\left(\alpha\right).\]
Finally,   adding the effects of the boundary terms, and
disregarding terms of higher orders we obtain
\begin{equation}
\alpha_{t}+\frac{\alpha_{y}^{2}}{2}\left(1+\frac{\Theta-1}
{2}\right)-\frac{\Theta-1}{2}\left\langle \frac{\alpha_{y}^{2}}{2}
\right\rangle =\frac{\Theta-1}{2}\left(1-
\frac{\Theta-1}{2}\right)I\left(\alpha\right).
\label{eq:Sivashinsky-Clavin}
\end{equation}
This  is the Sivashinsky-Clavin equation, written in units of
laminar flame speed (relative to the fresh gases). The lateral
mean value term comes from the boundary velocity and was absent in
the original article. It was later added by Joulin, who used it in
a number of papers, the first one being probably
\cite{joulincambray} and called it a counter-term. In
Sivashinsky-Clavin units the value
$\gamma=\frac{\Theta-1}{\Theta}$ was used to characterize gas
expansion and  all velocities were scaled by the laminar flame
speed relative to burnt gases $U_{b}=\Theta
U_{f}=U_{f}/(1-\gamma)$. In such units   equation
(\ref{eq:Sivashinsky-Clavin}) would be written as
\[
\alpha_{t}+\frac{\alpha_{y}^{2}}{2}\left(1-\frac{\gamma}{2}\right)
-\frac{\gamma}{2}\left\langle
\frac{\alpha_{y}^{2}}{2}\right\rangle
=\frac{\gamma}{2}\left(1-\frac{\gamma}{2}\right)I\left(\alpha\right).\]
In units of laminar flame speed relative to fresh gases $U_{f}$,
but using $\gamma$ as a parameter the equation takes the form
\[
\alpha_{t}+\frac{\alpha_{y}^{2}}{2}\left(1+\frac{\gamma}{2}\right)
-\frac{\gamma}{2}\left\langle
\frac{\alpha_{y}^{2}}{2}\right\rangle
=\frac{\gamma}{2}\left(1+\frac{\gamma}{2}\right)I\left(\alpha\right).\]

\section{Do we have the first or second time-derivative in the flame equations?\label{sec:Time-derivative}}

One more question concerns the order of time-derivative in the
Sivashinsky-Clavin equation, which also affects the structure of
nonlinear terms in the equation. The original DL dispersion
relation \cite{LandauLifFluid} is of the second order describing
two independent linear modes of the flame front perturbation: one
mode is growing and one is decaying. Unlike this, the Sivashinsky
equation  is of the first order in time capturing  the growing
mode only;  the Sivashinsky-Clavin equation has the same property.
When investigating development of the DL instability, the growing
mode dominates over the decaying one and the first-order
derivative looks quite sufficient. However, the second-order time
derivative proved to be of principal importance in many adjacent
problems of flame dynamics: propagation of tulip flames
\cite{DoldJoulin}, flame interaction with sound
\cite{Searby_and_Rochwerger_1991,Bychkov99}, with shocks
\cite{shocks-98} and with external turbulence
\cite{SearbyClavin86,AldredgeWilliams91,AkkermanBychkov03}. In all
these cases some external force redistributes energy between
growing and decaying modes, and it is incorrect to exclude the
decaying mode out of consideration. Besides, as we said above,
changing the order of the time derivative we also modify the
nonlinear terms of the equation. At this point, following comments
of one of the referees, we would like to add that, in general, even a 
first-order time derivative may produce a complicated spectrum with many 
stable and unstable modes. However, in the particular case of the linear DL 
instability for an infinitely thin planar flame front and a fixed wave number 
of perturbations the number of modes is unambiguously related to the order of time
derivative. The first order time-derivative of the Sivashinsky equation provides only
one mode (growing), while the original DL dispersion relation with the 
second order derivative has two modes (one is growing and one is decaying).

    Let us consider how the Sivashinsky-Clavin equation should be
modified to take into account the second time derivative. Within
the accuracy of the Sivashinsky-Clavin approach we can substitute
(\ref{eq:Sivashinsky}) into the second-order terms of
(\ref{eq:Sivashinsky-Clavin}) and find
\begin{equation}
\frac{\Theta-1}{2} \alpha_{t}+\alpha_{t}+
\frac{\Theta}{2}\alpha_{y}^{2}+ \left\langle
\frac{\alpha_{y}^{2}}{2}\right\rangle
 =\frac{\Theta-1}{2}I(\alpha)  \label{eq:19}
\end{equation}
or
\begin{equation}
I^{-1}\left(\alpha_{tt}\right)+\alpha_{t}+
I^{-1}\left(\frac{\partial}{\partial
t}\frac{\alpha_{y}^{2}}{2}\right)+ \frac{\Theta}{2}\alpha_{y}^{2}-
\frac{\Theta-1}{2}\left\langle \alpha_{y}^{2}\right\rangle
 =\frac{\Theta-1}{2}I(\alpha).  \label{eq:20}
\end{equation}
The operator $I^{-1}$ has a meaning of an integral, and it is
defined with the accuracy of a constant. Still, this constant is
included already into the counter-terms. One more trouble with
this operator concerns $I^{-1}$ acting on a constant. However, in
the case of equation (\ref{eq:20}), a constant under $I^{-1}$
would imply physically meaningless solutions like a planar
accelerating flame front, which may be obviously ruled out. One
can easily see that the linear terms of equation (\ref{eq:20})
coincide with the DL dispersion relation written in the limit of
$\Theta - 1 \ll 1$
\begin{equation}
\frac{\Theta+1}{2\Theta}I^{-1}\left(\alpha_{tt}\right)+\alpha_{t}
 =\frac{\Theta-1}{2}I(\alpha).  \label{eq:DL-linear}
\end{equation}
Equation (\ref{eq:20}) contains only one counter-term, because the
time-dependent nonlinear term involves the complete time
derivative and gives zero after averaging. Though equations
(\ref{eq:Sivashinsky-Clavin}) and (\ref{eq:20}) are mathematically
equivalent within the expansion in $\Theta -1\ll 1$, they may lead
to somewhat different conclusions about properties of curved
flames. For example, one of the most important questions in the
nonlinear theory of the DL instability is the velocity increase of
curved stationary flames. Let us consider propagation of such a
flame $\alpha(t,y)=-\Omega t + \alpha(y)$. In that case equations
(\ref{eq:Sivashinsky-Clavin}) and (\ref{eq:20})  reduce to
\begin{equation}
-\Omega+\frac{\alpha_{y}^{2}}{2}\left(1+\frac{\Theta-1}
{2}\right)-\frac{\Theta-1}{2}\left\langle \frac{\alpha_{y}^{2}}{2}
\right\rangle =\frac{\Theta-1}{2}\left(1-
\frac{\Theta-1}{2}\right)I\left(\alpha\right)
\label{eq:Siv-Clav-stat}
\end{equation}
and
\begin{equation}
-\Omega+ \frac{\Theta}{2}\alpha_{y}^{2}-
\frac{\Theta-1}{2}\left\langle \alpha_{y}^{2}\right\rangle
 =\frac{\Theta-1}{2}I(\alpha).  \label{eq:20-stat}
\end{equation}
As we can see, equations (\ref{eq:Siv-Clav-stat}) and
(\ref{eq:20-stat}) are different. It is interesting that equation
(\ref{eq:20-stat}) is consistent with the stationary theory
\cite{Bychkov98} within the accuracy of $(\Theta
-1)\alpha_{y}^{2}$, while (\ref{eq:Siv-Clav-stat}) is not. The
above calculations illustrate the fact that any rigorous expansion
in power series leaves plenty of freedom for mathematical
manipulations, which may lead to ambiguous physical conclusions.
Unfortunately, almost always people use expansion in power series
of a small parameter to obtain physical results beyond the
validity limits of the expansion. As a simple illustration,
suppose that we  calculated some value $a$ using expansion in
power series of $\varepsilon \ll 1$, and found $a=b$ in the zero
order approximation with $a=b(1-c\varepsilon)$ for the first
order. The same expression may be written in an infinite number of
equivalent mathematical forms like $a=b(1-[c+b-a]\varepsilon)$,
$a=b(1-2c\varepsilon)^{1/2}$, $a=b/(1+c\varepsilon)$, etc. Though
these forms are equivalent within the first order in $\varepsilon
\ll 1$, we come to different conclusions when investigating zero
points of $a$ with the help of these expressions. In the above
four versions of the same formula we find $a=0$ at
$\varepsilon=1/c$, $\varepsilon=1/(c+b)$, $\varepsilon=1/2c$ and
$\varepsilon=\infty$, respectively. One encounters a similar
trouble within both linear and nonlinear theories of the DL
instability. For example, the linear theories \cite{PelceClavin82}
and \cite{MatalonMatkowsky82} lead to noticeably different
expressions for the cut off wavelength of the DL instability,
though both theories are mathematically correct and equivalent
within the same accuracy of small wave numbers. Performing
manipulations similar to those described above, we can actually
obtain infinite number of absolutely different formulas for the
cut off wavelength in scope of the theory \cite{PelceClavin82}
keeping the same accuracy. So in the case of the linear DL instability 
the simple example $a=b(1-c\varepsilon)$ is sufficient to explain the discrepancy
between the analytical values for the cut-off wavenumber obtained by different
authors, as this formula is a simplified version of the dispersion relations obtained
in the two previously mentioned articles.

What does it mean if one tries to derive a non linear equation for the DL instability 
using perturbation methods ? In that case, as pointed out by one of the referees,
the perturbations concern operators instead of functions, which is a much subtler
subject. Of course, in this case, the example with $a=b(1-c\varepsilon)$ is ultimately
simplified, still it gives a rough idea why the non linear approaches like 
\cite{ZhdanovTrubnikov,Joulin91,Bychkov98,kazakovcst2002,kazakovpof2002} may lead to
different equations for a flame front even if all mathematical calculations are
performed correctly. Singular perturbation methods for partial differential 
operators are much more
sophisticated than for functions, and can involve a large number of new phenomena,
boundary layers being of course the most well-known example. However singular layers
can also occur during the time evolution, for instance initial layers for times close
to the initial conditions. See below for a discussion of the difficulties that could
happen in formal manipulations of second order in time equations.

 When one uses
expansion in powers of a small parameter, the nonlinear equation
for a flame front may be presented in an infinite number of forms. For example,
suppose that we have derived a time-dependent nonlinear equation
within the approach of weak nonlinearity as it was done in 
\cite{ZhdanovTrubnikov,kazakovcst2002}. Within the same accuracy
of calculations one can take square of the DL dispersion relation
(\ref{eq:DL-linear}) with any coefficient and add it to the
equation obtained. However, when we use the new version of the
nonlinear equation to study curved stationary flames, square of
the right-hand side of (\ref{eq:DL-linear}) makes a non-zero
contribution, while square of the left-hand side becomes zero (the
first term gives zero exactly, and the second term provides
nonlinearity of the fourth-order). Making such manipulations one
comes to a stationary equation like
\begin{equation}
-\Omega+
\left(\frac{1}{2}+C_{1}\right)\alpha_{y}^{2}+C_{2}\left[I(\alpha)\right]^{2}-
C_{1}\left\langle \alpha_{y}^{2}\right\rangle - C_{2}\left\langle
\left[I(\alpha)\right]^{2}\right\rangle
 =\frac{\Theta-1}{2}I(\alpha).  \label{eq:stat-any}
\end{equation}
with almost arbitrary coefficients $C_{1}$ and $C_{2}$. The only
restriction on the factors $C_{1}$ and $C_{2}$ is that they should
tend to zero sufficiently fast as $\Theta \rightarrow 1$, and that
the development in $\Theta-1$ of the stationary solution is the
same as the original (\ref{eq:20-stat}). It gives
$C_{1}=(\Theta-1)/2+O(\Theta-1)^{2}$ and $C_{2}=O(\Theta-1)^{2}$
for small $(\Theta-1)$, but it must be admitted that these
restrictions are
 too loose taking into account realistically large values of $\Theta=5-8$
 (also, it must be remembered that curvature-related terms have to be added to (\ref{eq:stat-any}) in
agreement with the linear theory of the DL instability).
 This result is rather discouraging, because it
leaves no hope to obtain an unambiguous formula for the flame
velocity by using perturbation theories. As an illustration of
this fact, the authors of the two companion papers
\cite{kazakovcst2002} and \cite{kazakovpof2002} have produced two
different formulas for the flame velocity. The first-order
approximation in \cite{ZhdanovTrubnikov,kazakovcst2002,boury},
contrary to the reasoning used above to obtain equation
(\ref{eq:stat-any}), employed the DL-dispersion relation for the
growing mode only
\begin{equation}
\frac{\partial \alpha}{ \partial t} = \frac{\Theta}{\Theta + 1}
\left( \left[\Theta +1-1/\Theta \right]^{1/2}-1 \right)
I\left(\alpha\right). \label{eq:DL-linear-growing}
\end{equation}
Using (\ref{eq:DL-linear-growing}) within the second-order
approximation one can always convert time-dependent nonlinear
terms into time-independent terms and vice versa. We would like to stress
 that manipulations with the time-derivatives is not something 
that we have invented in the present paper. They have been performed already 
in a number of papers on flame dynamics, leading to ambiguous physical results. 
In the present paper, we just clarify the ambiguity and point out the danger of such
manipulations.

At this point one of the referees asks, if it is not safer to transform second-order
time derivatives into space-derivatives. Unfortunately, even in that case some
ambiguity remains in the non linear terms. Besides, as we pointed out above, by
getting rid of the second-order derivatives one loses the possibility to study a large
number of effects like tulip flames, flame interaction with shocks and many other phenomena. At present, we
do not know how to avoid the ambiguity in the perturbation theory of the non linear
equation for a flame front. In the present section we rather formulate a question than
give an answer. We hope to insist here on the fact that an infinite number of
different non linear equations for a flame front can be obtained by high order formal 
perturbation methods. Among these formally equivalent equations, some will give bad
 quantitative results (particularly if the development parameter is not small, 
 which is unfortunately the case for flame fronts). 
 Some will even give bad qualitative results, 
 for instance we could have second order in time equations which, without forcing, do
 not approach a first order in time dynamics (which is well known to attract the
 dynamics for a flame without forcing). Even worse, we could have equations with
  pathological mathematical properties 
  (this is particularly possible with time derivatives in the non linear terms). 
  Although in the rest of the paper, we insist on the quantitative agreement 
  on the flame velocity, it must be kept in mind that at
  some point, the time evolution of the proposed models must also be compared 
  with direct numerical simulations and found satisfactory. 

On the other
hand, without using perturbation methods one comes to a rather
complex set of equations \cite{bychkovzaytsevakkerman} with almost
zero hope to solve it analytically, and also very difficult to solve numerically. 
Luckily, in the case of curved
stationary flames the problem of flame velocity has been solved
with the help of
 direct numerical simulations \cite{Bychkov-et.al-96}; later
calculations \cite{Kadowaki99,Travnikov-et.al-00} confirmed the
original results.

\begin{figure}

\begin{center}
\includegraphics[angle=-90,width=1.0\textwidth]{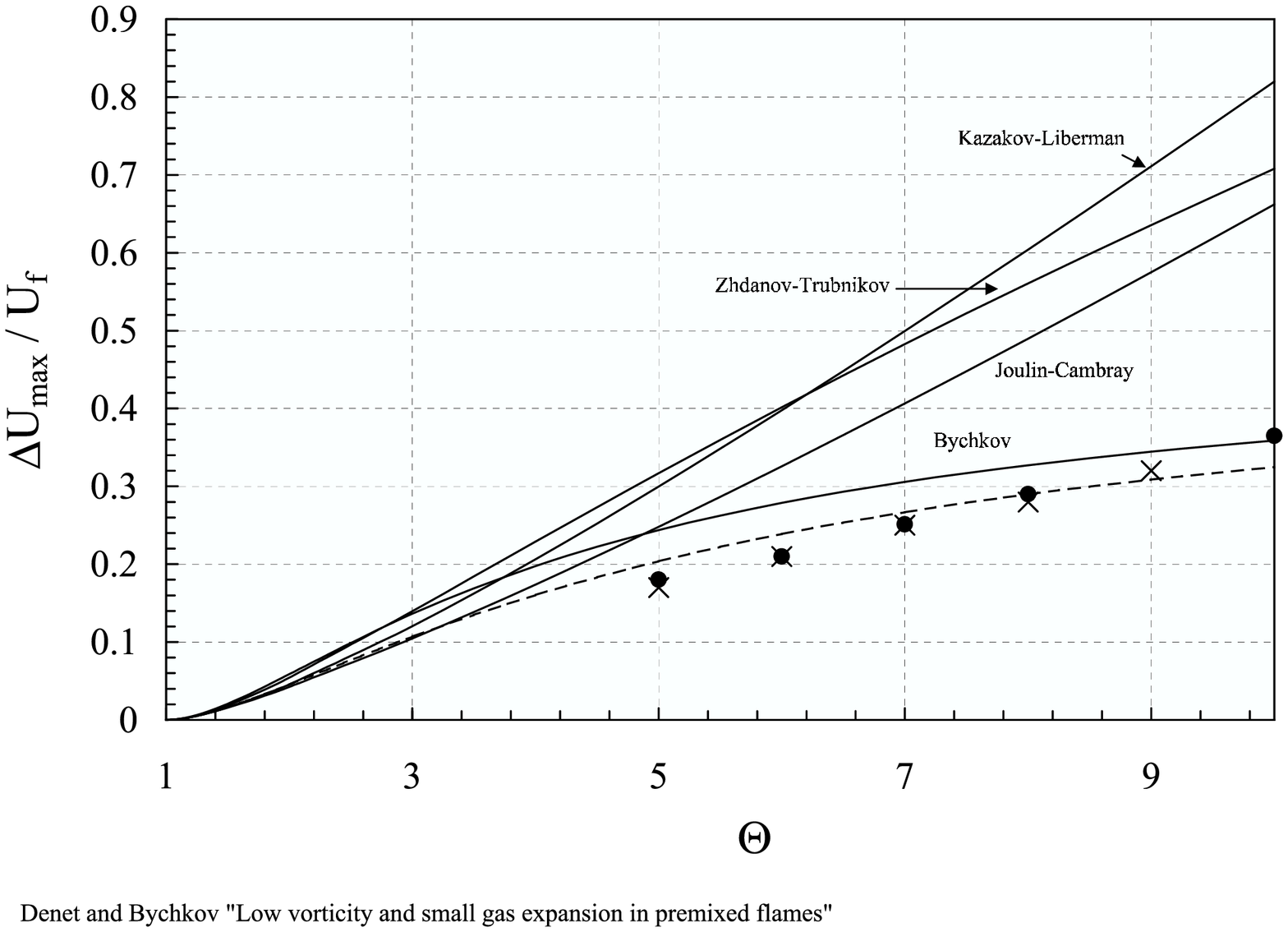}
\end{center}
\caption{Maximal velocity increase for curved stationary flames
scaled by the planar flame velocity versus the thermal expansion
$\Theta$. The markers show results of direct numerical simulations
\protect\cite{Bychkov-et.al-96,Travnikov-et.al-00} (circles) and
\protect\cite{Kadowaki99} (crosses). The solid lines correspond to
the analytical results of
\protect\cite{ZhdanovTrubnikov,Bychkov98,kazakovcst2002,joulincambray}.
The dashed line presents equation (\ref{eq:V-analyt2}).}
 \label{fig:V-max}
\end{figure}

The uncertainty in the rigorous perturbation nonlinear theories of
the DL instability increases the role of simple phenomenological
models like that proposed in \cite{joulincambray}. Using the
models one can obtain qualitative or even semi-quantitative
understanding of flame dynamics, which may be checked and
corrected quantitatively in direct numerical simulations. The
model \cite{joulincambray} included first-order time-derivative
similar to the Sivashinsky equation (\ref{eq:Sivashinsky}). When
second-order derivatives are important, similar model can be
constructed on the basis of equation (\ref{eq:20}). Comparing
equations (\ref{eq:20}) and (\ref{eq:DL-linear}) we can easily
extrapolate (\ref{eq:20}) to the case of realistic $\Theta$ as
\begin{equation}
\frac{\Theta+1}{2\Theta}I^{-1}\left(\alpha_{tt}\right)+\alpha_{t}
+I^{-1}\left(\frac{\partial}{\partial
t}\frac{\alpha_{y}^{2}}{2}\right)+\left(\frac{1}{2}+C_{1}\right)\alpha_{y}^{2}-
C_{1}\left\langle \alpha_{y}^{2}\right\rangle
 =\frac{\Theta-1}{2}I(\alpha).  \label{eq:new-model}
\end{equation}
Similar equation was proposed in Boury's thesis \cite{boury},
in the spirit of the phenomenological theory used in
\cite{DoldJoulin} to study tulip flames. The linear part of
 (\ref{eq:new-model}) is a well-known DL
dispersion relation.
Let us note however that including
curvature effects in this type of equation may be non trivial, because actually
the Markstein lengths are frequency-dependent
(see \cite{joulinfreqmark,clavinjoulinfreqmark,denettoma}).
The unknown coefficient
$C_{1}$ may be adjusted by using direct numerical simulations of
curved stationary flames. In \cite{boury} the coefficient was
chosen to provide the same stationary amplitude as either a third order
gas expansion theory or direct numerical simulations.
 However, we believe that fitting  direct
numerical simulations is a better idea. As explained before, a
high order perturbation theory can be written in several
equivalent ways, with different quantitative results for large
$\Theta$ (see for instance, equations (\ref{eq:Siv-Clav-stat}) and
(\ref{eq:20-stat})). Below, we illustrate that this kind of fit
can actually be used to obtain almost any curved flame velocity,
with only some restrictions for small gas expansion.
The stationary version of
(\ref{eq:new-model}) is
\begin{equation}
-\Omega +\left(\frac{1}{2}+C_{1}\right)\alpha_{y}^{2}-
C_{1}\left\langle \alpha_{y}^{2}\right\rangle
 =\frac{\Theta-1}{2}I(\alpha).  \label{eq:new-model-stat}
\end{equation}
Equation (\ref{eq:new-model-stat}) may be solved analytically
\cite{Thual-et.al-85,joulincambray}, which leads to the maximal
velocity increase
\begin{equation}
\Omega_{max}=\frac{(\Theta -1)^{2}}{8\left(1+2C_{1}\right)^{2}}.
\label{eq:V-stat}
\end{equation}
The maximal velocity increase obtained in direct numerical
simulations \cite{Bychkov-et.al-96,Kadowaki99,Travnikov-et.al-00}
is plotted in figure \ref{fig:V-max} by markers. The curves of
figure \ref{fig:V-max} show the analytical formulas for the
velocity increase, which follow from the theories
\cite{ZhdanovTrubnikov,Bychkov98,kazakovcst2002,joulincambray}.
As we can see, the formulas
\cite{ZhdanovTrubnikov,kazakovcst2002,joulincambray}
overestimate the velocity increase noticeably, especially for
$\Theta = 8$ corresponding to stoichiometric methane and propane
flames. So far, the formula proposed in \cite{Bychkov98}
\begin{equation}
\Omega_{max}=\frac{\Theta}{2}\frac{(\Theta
-1)^{2}}{\Theta^{3}+\Theta^{2}+3\Theta-1} \label{eq:V-analyt}
\end{equation}
provides the best analytical fit for the numerical results. At
this point we have to note that, as was remarked in \cite{boury},
neither of the papers
\cite{ZhdanovTrubnikov,Bychkov98,kazakovcst2002,kazakovpof2002}
included Joulin counter-terms (see again \cite{joulincambray}),
which would lead to considerable quantitative corrections to all
these results including equation (\ref{eq:V-analyt}). However,
one can obtain almost any velocity increase in
scope of the perturbation approaches, and the formula
(\ref{eq:V-analyt}) of \cite{Bychkov98} as well as other formulas
of \cite{ZhdanovTrubnikov,kazakovcst2002,kazakovpof2002} may be
equally treated as analytical guesses rather than unambiguous
results.
 Taking (\ref{eq:V-analyt})
as the estimate for the velocity increase we find
\begin{equation}
C_{1}=\frac{(\Theta
-1)}{4(2\Omega_{max})^{1/2}}-\frac{1}{2}=\frac{1}{4}\left(\Theta^{2}+\Theta+3-1/\Theta\right)^{1/2}-\frac{1}{2}.
\label{eq:C-1}
\end{equation}
The model equation (\ref{eq:new-model}) involves also
time-dependent nonlinear terms, which cannot be adjusted with the
help of direct numerical simulations for stationary flames.
However, studies of curved flame stability
\cite{Bychkov-et.al-99,PetchenkoBychkov} indicate that
time-dependent nonlinear terms are of minor importance. For
simplicity, when constructing a qualitative model the
time-dependent nonlinear term may be omitted. Still, there is
another problem with formula (\ref{eq:V-analyt}), namely, equation
(\ref{eq:C-1}) does not reproduce correct asymptotics for $C_{1}$
at $\Theta\rightarrow 1$. Making slight modifications of
(\ref{eq:V-analyt}) we can remedy this trouble, since, as we have
pointed out above, the stationary flame velocity is almost a free
parameter in the nonlinear perturbation theories. For example, we
can choose
\begin{equation}
\Omega_{max}=\frac{\Theta}{2}\frac{(\Theta
-1)^{2}}{\Theta^{3}+2\Theta^{2}+5\Theta-4} \label{eq:V-analyt2}
\end{equation}
with respective corrections to the coefficient $C_{1}$
\begin{equation}
C_{1}=\frac{(\Theta
-1)}{4(2\Omega_{max})^{1/2}}-\frac{1}{2}=\frac{1}{4}\left(\Theta^{2}+2\Theta+5-4/\Theta\right)^{1/2}-\frac{1}{2}.
\label{eq:C-1-2}
\end{equation}
The velocity increase (\ref{eq:V-analyt2}) is shown in figure
\ref{fig:V-max} by the dashed line. As we can see, it provides
even better agreement with direct numerical simulations than
(\ref{eq:V-analyt}). 

Of course, the above calculations is not the way to construct an unambiguous rigorous
equation to calculate the flame propagation velocity. Particularly, one of the referees
points out that the second order time derivative is useless, if the ultimate goal is
only to compute the flame velocity. But we recall that the non linear theory of the DL
instability in an inevitable starting point for many other problems like oblique
flames, tulip flames, flame interaction with turbulence, with acoustics or shock
waves, burning in tubes with heat losses. As an example, the non linear equations 
(see for instance \cite{joulincambray,Bychkov98}) developed to describe
 the DL instability were later
used with some modifications to study turbulent flames in 
\cite{cambrayjoulin,denetfrankel,Bychkov2000,Zaytsevbychkov2002,AkkermanBychkov03}.
In the same way, equation ((\ref{eq:new-model})) can also be modified 
to get an understanding of other, much more complicated phenomena.

\section{Conclusion\label{sec:Conclusion}}

In this article, starting from a low vorticity approach
\cite{bychkovzaytsevakkerman} proposed to describe premixed flames
 in a coordinate-free way, we have developed this formulation
in powers of the gas expansion parameter. It appears that at the
lowest order in gas expansion, the Frankel equation (equivalent of
the Sivashinsky equation for the coordinate-free case) is
recovered. At the second order, we have shown that complications
arise in the case of oblique flames. On the contrary, if the
tangential velocity is small (planar on average and expanding
flames), we have obtained a modified form of the Frankel equation,
with a correction of the surface charge. In the planar case, we
have shown that this modified equation reduces to the
Sivashinsky-Clavin equation (a second order in gas expansion
correction to the original Sivashinsky equation). We have thus
shown that the small vorticity formulation contains  the
Sivashinsky-Clavin equation as a particular case. A direct
numerical solution of this formulation, although difficult, could
describe both the slow dynamics of a flame without external
forcing, and the rapid evolution that takes place under some
conditions (acoustic forcing, interaction with shock waves). On
the contrary, the equations obtained here as the lowest orders of
a development in gas expansion are inherently limited to a slow,
first order in time dynamics. Although potentially easier to solve
than the full small vorticity equations (at least without
tangential blowing) it must be recalled that in the planar case,
in order to obtain good quantitative agreement with numerical
simulations, Joulin and Cambray \cite{joulincambray} have been
obliged to perform some empirical modifications of the
coefficients of the Sivashinsky-Clavin equation. The purpose of
these new coefficients was to  describe better the instability
growth rates and the amplitude of stationary cellular flames. We
suggest different ways to construct similar phenomenological
equations, particularly taking into account second-order time
derivatives inherent to the DL dispersion relation. We also
discuss the best way of adjusting the numerical coefficients of
the model equation using recent results of direct numerical
simulations for the velocity increase because of the DL
instability \cite{Bychkov-et.al-96,Kadowaki99,Travnikov-et.al-00}.
It remains to be seen if the same type of modification has to be
used in the coordinate-free case, both for the small vorticity
equations and its small gas expansion, low frequency limit. In any
event, we hope that the present article has served to explain the
relations between different existing approaches to the problem of
nonlinear premixed flames dynamics.


\emph{Acknowledgments} : B. Denet would like to thank P. Clavin
for discussions on the history of the Sivashinsky-Clavin article,
and G. Joulin for explanations on the Joulin-Cambray article,
particularly on the role of the counter-terms. This work has been
supported in part by the Swedish Research Council (VR).

\bibliography{turb2d.reflib}
\newpage

\end{document}